\documentclass[11pt]{article}
\usepackage{amsmath,amsfonts,amssymb,bm}
\RequirePackage[a4paper,body={16cm,26cm}]{geometry}
\usepackage{pslatex}
\usepackage{graphicx,color}
\usepackage{etex}
\usepackage[T1]{fontenc}
\usepackage[utf8]{inputenc}
\usepackage[english]{babel}
\usepackage[noindentafter,pagestyles]{titlesec}
\usepackage{textcomp}
\usepackage{dcolumn}
\usepackage{amsmath}
\usepackage{txfonts}
\usepackage{fancyhdr}
\usepackage{threeparttable}
\usepackage{xspace}
\usepackage[superscript]{cite}
\usepackage{subcaption}
\usepackage{pictex}

\usepackage[algo2e,ruled,lined,titlenumbered,commentsnumbered]{algorithm2e}
\usepackage[export]{adjustbox}
\usepackage{graphicx,color}
\usepackage{algpseudocode}
\usepackage{algorithm2e}
\usepackage{algorithm}
\usepackage{authblk}
\usepackage{hyperref}
\usepackage{float}
\newcommand{\address}[1]{\def\@address{#1}}
\title{Asynchronous scalable version \\of the Global-Local non-invasive coupling}
\author{Ahmed EL KERIM$^{1,3}$, Pierre GOSSELET$^2$, Frédéric MAGOUL\`ES$^3$, \\}
\date{$^1$ Universit\'e Paris-Saclay, ENS Paris-Saclay, CNRS, LMT, Gif-sur-Yvette, France, ahmed.elkerim@ens-paris-saclay.fr\\
	$^2$ Université de Lille, CNRS, Centrale Lille / LaMcube, pierre.gosselet@univ-lille.fr\\
	$^3$ Universit\'e Paris-Saclay, CentraleSup\'elec / MICS , Gif-sur-Yvette, France, frederic.magoules@hotmail.com\\
}
\begin{document}
	\maketitle
	\newcommand{\dom}{\ensuremath{\Omega}}
	\newcommand{\inter}{\ensuremath{\Gamma}}
	
	\newcommand{\F}{\ensuremath{^F}}
	\newcommand{\G}{\ensuremath{^G}}
	\newcommand{\GT}{\ensuremath{^{G^T}}}
	\newcommand{\Fs}{\ensuremath{^{(s),F}}}
	\newcommand{\Gs}{\ensuremath{^{(s),G}}}
	\newcommand{\Gz}{\ensuremath{^{(0),G}}}

	\newcommand{\FsT}{\ensuremath{^{(s),F^T}}}
	\newcommand{\GsT}{\ensuremath{^{(s),G^T}}}
	\newcommand{\GzT}{\ensuremath{^{(0),G^T}}}
	
	\newcommand{\s}{\ensuremath{^{(s)}}}
	\newcommand{\z}{\ensuremath{^{(0)}}}
	\newcommand{\sT}{\ensuremath{^{(s)^T}}}
	\newcommand{\zT}{\ensuremath{^{(0)^T}}}
	
	\newcommand{\R}{\ensuremath{^R}}
	\newcommand{\Gm}{\ensuremath{^{G^{-1}}}}

	\newcommand{\bK}{\ensuremath{\mathbf{K}}}
	\newcommand{\bA}{\ensuremath{\mathbf{A}}}
	\newcommand{\bI}{\ensuremath{\mathbf{I}}}
	\newcommand{\bJ}{\ensuremath{\mathbf{J}}}
	\newcommand{\bS}{\ensuremath{\mathbf{S}}}
	\newcommand{\bT}{\ensuremath{\mathbf{T}}}
	\newcommand{\bP}{\ensuremath{\mathbf{P}}}
	\newcommand{\bM}{\ensuremath{\mathbf{M}}}
	
	\newcommand{\bSnl}{\ensuremath{\mathcal{S}}}
	
	\newcommand{\bu}{\ensuremath{\mathbf{u}}}
	\newcommand{\bp}{\ensuremath{\mathbf{p}}}
	\newcommand{\bq}{\ensuremath{\mathbf{q}}}
	\newcommand{\bv}{\ensuremath{\mathbf{v}}}
	\newcommand{\br}{\ensuremath{\mathbf{r}}}
	\newcommand{\bb}{\ensuremath{\mathbf{b}}}
	\newcommand{\bx}{\ensuremath{\mathbf{x}}}
	
	\newcommand{\f}{\ensuremath{\mathbf{f}}}
	\newcommand{\foint}{\ensuremath{\mathbf{f}_{int}}}
	\newcommand{\foext}{\ensuremath{\mathbf{f}_{ext}}}
	\newcommand{\foexti}{\ensuremath{\mathbf{f}_{ext,i}}}
	\newcommand{\foextb}{\ensuremath{\mathbf{f}_{ext,b}}}
	
	\newcommand{\lam}{\ensuremath{\boldsymbol{\lambda}}}
	
		\begin{abstract}
			 The Global-Local non-invasive coupling is an improvement of the submodeling technique, which permits to locally enhance structure computations by introducing patches with refined models and to take into accounts all the interactions. In order to circumvent its inherently limited computational performance, we propose and implement an asynchronous version of the method. The asynchronous coupling reduces the dependency on communications, failures, and load imbalance. We present the theory and the implementation of the method in the linear case and illustrate its performance on academic cases inspired by actual industrial problems.
		\end{abstract}
	\section{Introduction}
	The non-invasive global-local coupling is an iterative technique that aims at making accurate the well-known submodeling technique  \cite{kelley1982, ransom1992computational, cormier_Aggressive_1999}. It is strongly related to many reanalysis techniques \cite{jara-almonte_specified_1988, whitcomb_iterative_1991, whitcomb_application_1993} and domain decomposition methods \cite{HECHT:2009:NZS}.

	Starting from a global simplified model, this technique allows inserting local alterations (geometry, material, load and mesh) and evaluate their effect without heavy intervention on the initial model. The method is non-invasive in the sense that it is adapted to coupling commercial (closed) and research software, its first implementation in Abaqus was proposed in \cite{gendre2009non}.
	
	This philosophy was successfully applied in many contexts like: the introduction of local plasticity and geometrical refinements \cite{gendre2009non}, the computation of the propagation of cracks in a sound model \cite{duval2014non}, the evaluation of stochastic effects with deterministic computations \cite{safatly2012methode,nouy2017}, the taking into account of the exact geometry of connectors in an assembly of plates \cite{GUGUIN.2016.1}. In \cite{Duval2014} the method was used in order to implement a nonlinear domain decomposition method \cite{keyes1995aerodynamic,cresta2007nonlinear,hinojosa2014,NEGRELLO.2016.1} in a non-invasive manner in code\_aster. Extension of the approach to explicit dynamics was proposed in \cite{bettinotti2013coupling}, improved in \cite{bettinotti2014}  and applied to the prediction of delamination under impact loading in \cite{bettinotti2017}. 
	
	All the above applications were developed in a synchronous framework which has been taken advantage of by the use of accelerators (Aitken, quasi-Newton, Krylov), see \cite{blanchard.2018.1} were the method is proved to be an implementation of an alternating Dirichlet-Robin Schwarz domain decomposition method where the Robin parameter corresponds to the condensation of the coarse domain covered by the patch.Unfortunately, even if fast convergence is often observed, the method possesses inherent limitations in terms of computational performance.
	
	 The objective of this work is to propose an asynchronous version of the algorithm.  Asynchronous parallel computation, first introduced by \cite{Chazan}, was then the subject of several theoretical works to prove its convergence in linear and nonlinear situations \cite{Miellou, Baudet}. Recent works show that domain decomposition methods are well adapted to asynchronous parallel computation and may lead to fascinating results \cite{Magoules1, Garay, Gbikpi-Benissan2, Chau2}. The idea is to allow each processor to move at its speed without waiting for the others, only considering the latest version of the data available. 
	
	The paper is organized as follows: Section~2 gives the basics of the Global-Local algorithm, Section~3 presents its asynchronous version and Section~4 provides illustrations in the linear case. 
	
	\section{Global-Local coupling}
	
\begin{figure}[H]
		\null\hfill
		\begin{subfigure}[b]{0.3\linewidth}
			\centering
			\includegraphics[width=\textwidth]{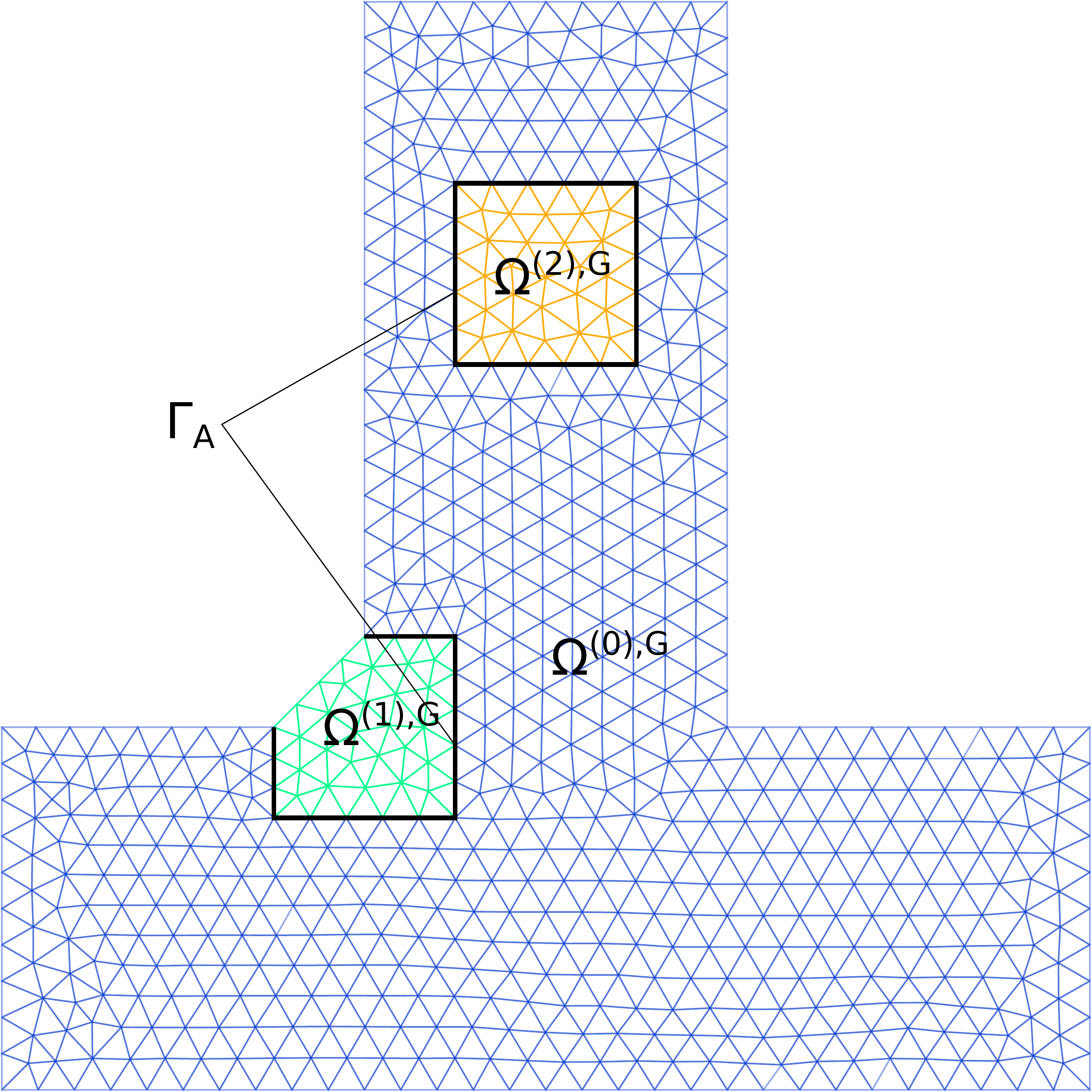}
			\caption{Global problem}\label{Global problem}
		\end{subfigure}
		\hfill
		\begin{subfigure}[b]{0.3\linewidth}
			\centering
			\includegraphics[width=.5\textwidth]{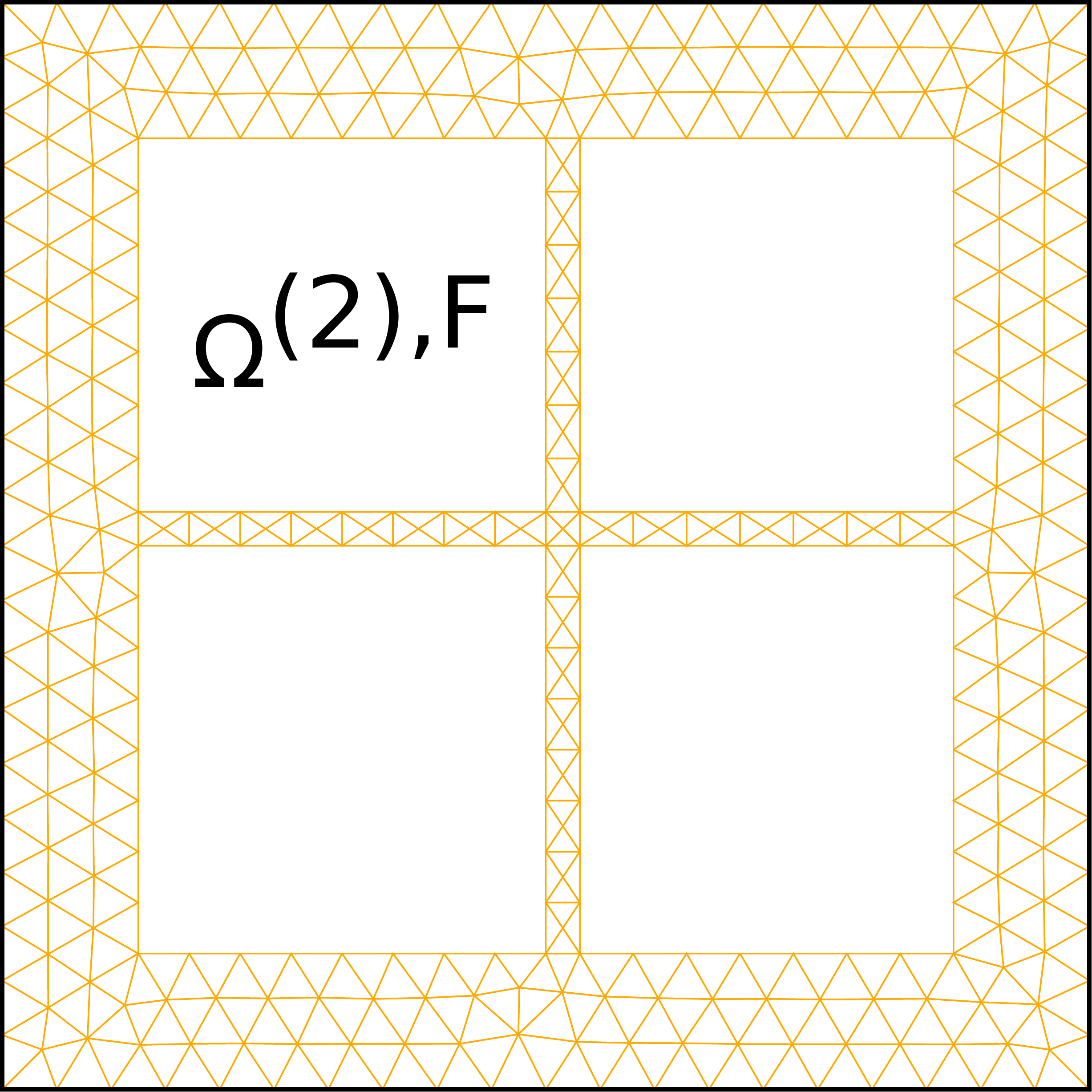}
			
			\null
			
			\includegraphics[width=.5\textwidth]{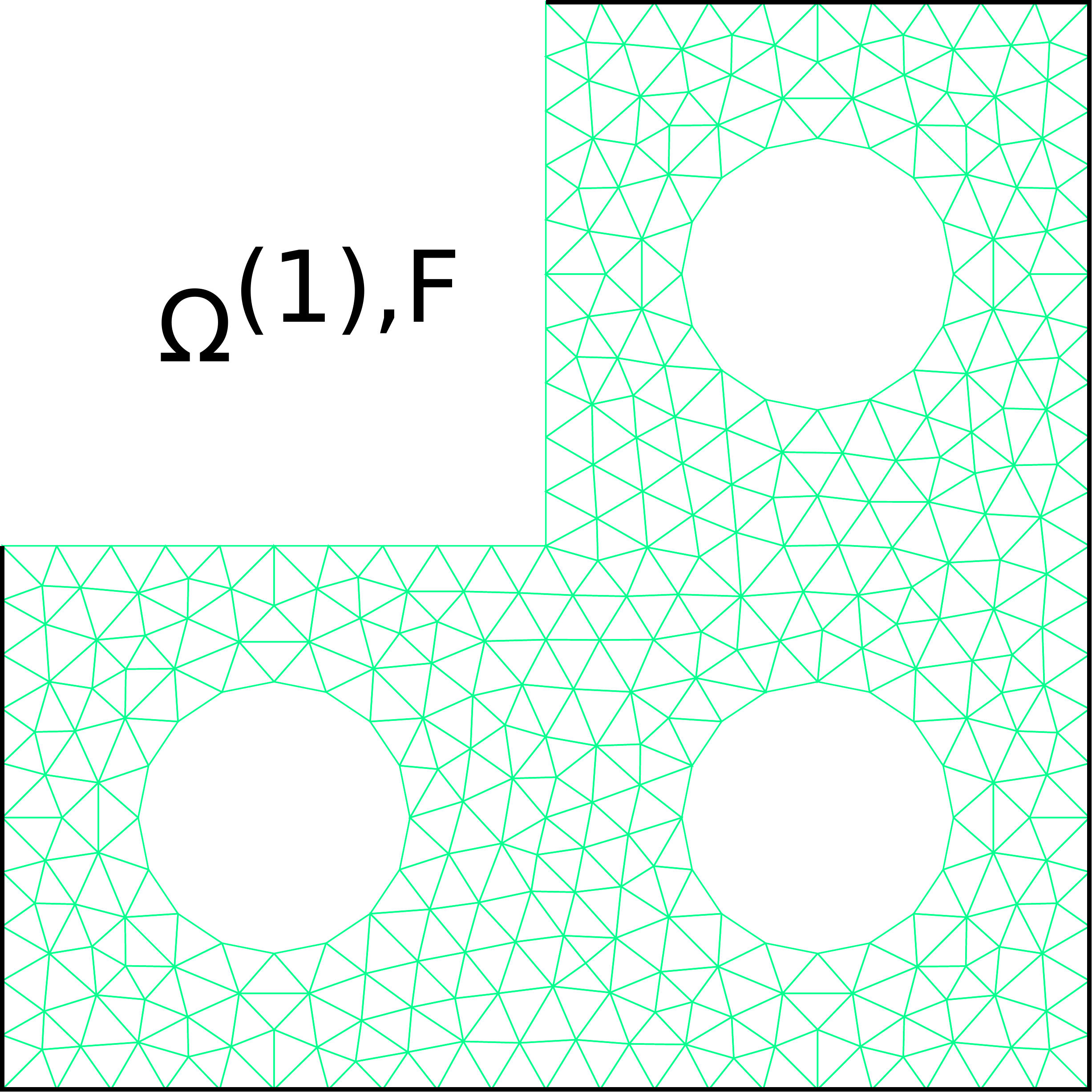}
			\caption{Refined zones of interest}     \label{Zones of interest}
		\end{subfigure}
		\begin{subfigure}[b]{0.3\linewidth}
			\centering
			\includegraphics[width=\textwidth]{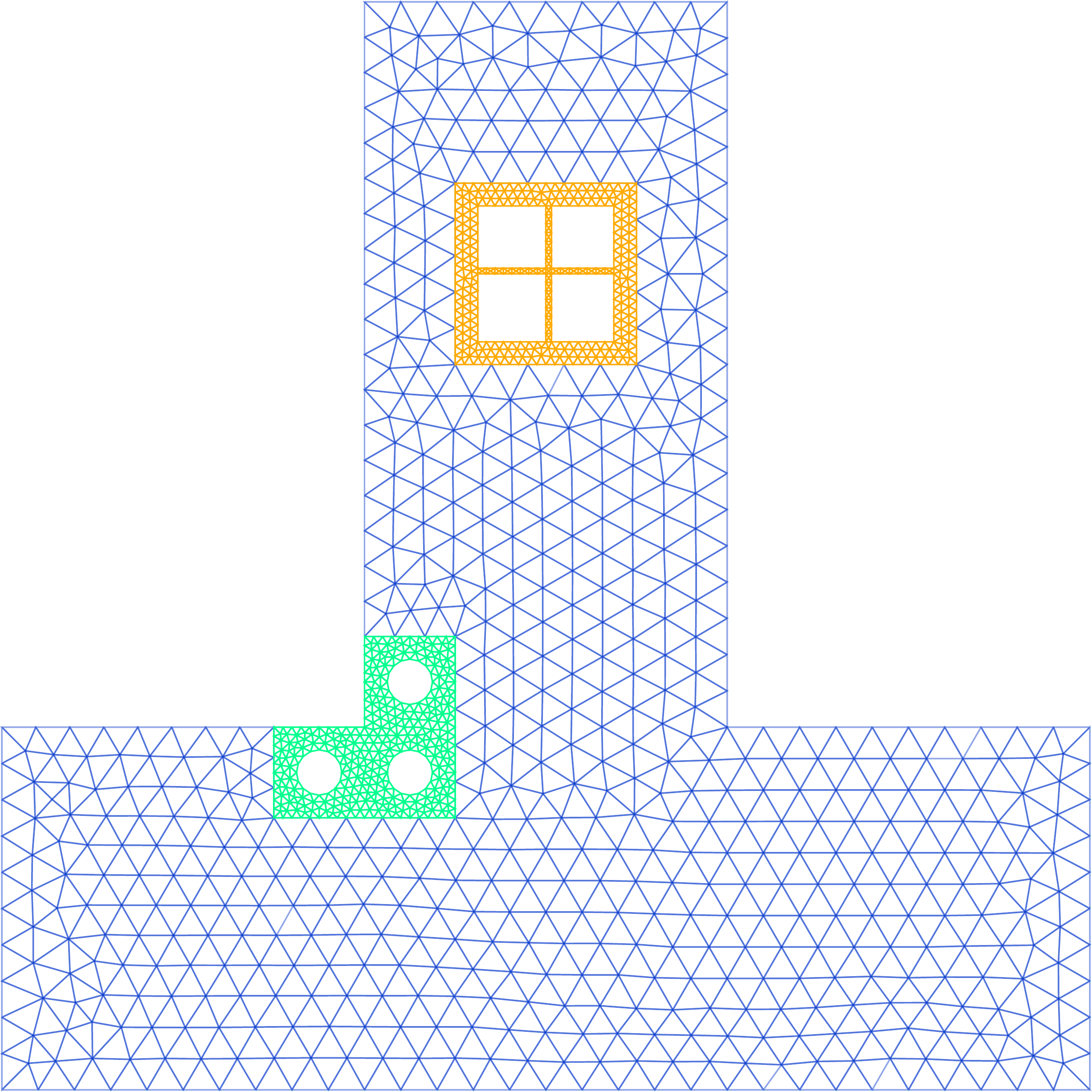}
			\caption{Reference problem}     \label{Reference problem}
		\end{subfigure}
		\hfill\null
		\caption{Models and subdomains for the Global/Local coupling}\label{fig:scenario}
	\end{figure}

	We consider linear elliptic equations, typically arising from static thermal or elastic (small deformations and small displacements) problems. As in the submodeling approach, the starting point of the Global-Local coupling is a Global (index $G$) model which is a simplification of the problem to be solved adapted to a fast calculation and able to give a correct representation of the long-distance fluxes (see Fig.~\ref{Global problem}). The Global model is insufficient in certain zones of interest (denoted by $\Omega^{(s)}$ with $s>0$) which need refined geometry, material law and adapted meshes (see Fig.~\ref{Zones of interest}). A local reanalysis is then conducted in the Fine models (index $(s),F$, $s>0$) with Dirichlet conditions inherited from the Global computation. The submodeling approach would stop at this point, but large errors could be obtained, in particular if the patches evolve nonlinearly, compared to the Reference computation (Fig~\ref{Reference problem}, index $R$) where the Fine zones of interest replace their Global counterparts (index $(s),G$, $s>0$) and a complete computation is run. Indeed, in the submodeling Local-to-Global and Local-to-other-Local interactions are totally omitted. 
	
	The error committed during the coupling can be measured by the lack of balance between the Fine zone of interest and their Global neighborhood (index $(0),G$). The Global-Local method consists in evaluating this residual and re-imposing it in the Global model as an extra load, in a Richardson iteration manner.
	
	For simplicity reason, we derive the method in after finite element discretization. We note $\bK$ the stiffness matrices (or their thermal equivalent), which are symmetrical positive semidefinite (definite as soon as enough Dirichlet condition is given), $\foext$ the generalized load vector, and $\bu$ the unknown vector (temperature or displacement). 
	
	The interface $\Gamma$ is defined as the boundary of patches: for each subdomain $\Gamma\Gs=\partial\Omega\Gs \setminus \partial\Omega\G$ and globally $\Gamma\G=\cup\Gamma\Gs$. The Fine version of the local interfaces $\Gamma\Fs=\partial\Omega\Fs\setminus\partial\Omega\R$ is geometrically conforming with $\Gamma\Gs$ but can be meshed more finely. We introduce the global trace operator $\bT\G:\Omega\G\to\Gamma\G$, and the local fine traces $\bT\Fs:\Omega\Fs\to\Gamma\Fs$, note that their transpose is the extension-by-zero operator. We note $\bJ\s: \Gamma\Gs\to\Gamma\Fs$ the interpolation operator between the meshes. Finally, $\bA\Gs:\Gamma\Gs\to\Gamma\G$ is the assembly operators which inject the subdomain's interface into the global interface.\medskip
	
	Let $\bp$ be a nodal effort defined on the interface $\Gamma$, initialized by 0, whose role will soon be made clear: the Global problem can be written as:
	\begin{equation} \label{eq:glo}
		\bK\G\bu\G = \foext\G + \bT\GT \bp,
	\end{equation}
	and the interface nodal reaction can be post-processed in the Global zone not covered by patches, on referred to as complement zone, and numbered $(0)$:
	\begin{equation}\label{eq:gloPost}
		\lam\Gz = \bT\Gz(\bK\Gz\bu\Gz - \foext\Gz)
	\end{equation}
	
	The Fine problems can be written as:
	\begin{equation}\label{eq:loc}
		\begin{pmatrix}
			\bK\Fs & \bT\FsT\\
			\bT\Fs & 0 \\
		\end{pmatrix}
		\begin{pmatrix}
			\bu\Fs\\
			-\lam\Fs\\
		\end{pmatrix}= 	\begin{pmatrix}
			\foext\Fs\\
			\bJ\s\bA\sT\bT\G\bu\G\\
		\end{pmatrix},
	\end{equation}
	$\lam\Fs$ is the Lagrange multiplier associated with the Dirichlet condition. The minus sign is used to make it interpretable as the nodal reaction exerted by the surrounding of the patch.
	
	It is then possible to evaluate the compatibility between the models as the lack of balance at the interface:
	\begin{equation}\label{eq:res}
		\br = -\left(\bA\z \lam\Gz + \sum_{s>0} \bA\s \bJ\sT\lam\Fs\right)
	\end{equation}
	If $\br$ is not small enough, it is injected in the computation in a modified Richardson iteration way: 
	\begin{equation} \label{eq:richard}
		\bp \leftarrow \bp + \omega\br
	\end{equation}
	where $\omega$ is a relaxation parameter to be determined.
	
	Under the chosen hypothesis, it can be proved that the method converges to the solution of the reference problem for a sufficiently small relaxation. In fact convergence is more general and can be improved by acceleration, in particular Aitken, see among others \cite{nouy2017, blanchard.2018.1}.
	
	The Global-Local coupling is recalled in Algorithm~\ref{alg:staSit}.
	
	\begin{algorithm2e}\caption{Non-invasive synchronous stationary iterations}\label{alg:staSit}\DontPrintSemicolon
		Arbitrary initialization $\bp_{0}\G$, Relaxation parameter $\omega$\;
		\For{$j\in\left[0,\cdots,m\right]$}{
			{\textit{Global} solve for $\bu\G_{j}$ for given $\bp_{j}$, eq.~\eqref{eq:glo}}  \;
			Postprocess $\lam\Gz_j$, eq.~\eqref{eq:gloPost}\;
			{Parallel \textit{fine} solve, for $s>0$ obtain $\bu\Fs_j$ and $\lam\Fs_j$ from imposed $\bT\G\bu\G$, eq.~\eqref{eq:loc}}\;
			{Assemble residual: $\br_{j}$, eq.~\eqref{eq:res}}\;
			{Update: $\bp_{j+1}=\bp_{j}+\omega\br_{j}$}}
	\end{algorithm2e}

\begin{figure}[H]
		\null\hfill
		\begin{subfigure}{.45\textwidth}\centering
			\includegraphics[width=0.6\textwidth]{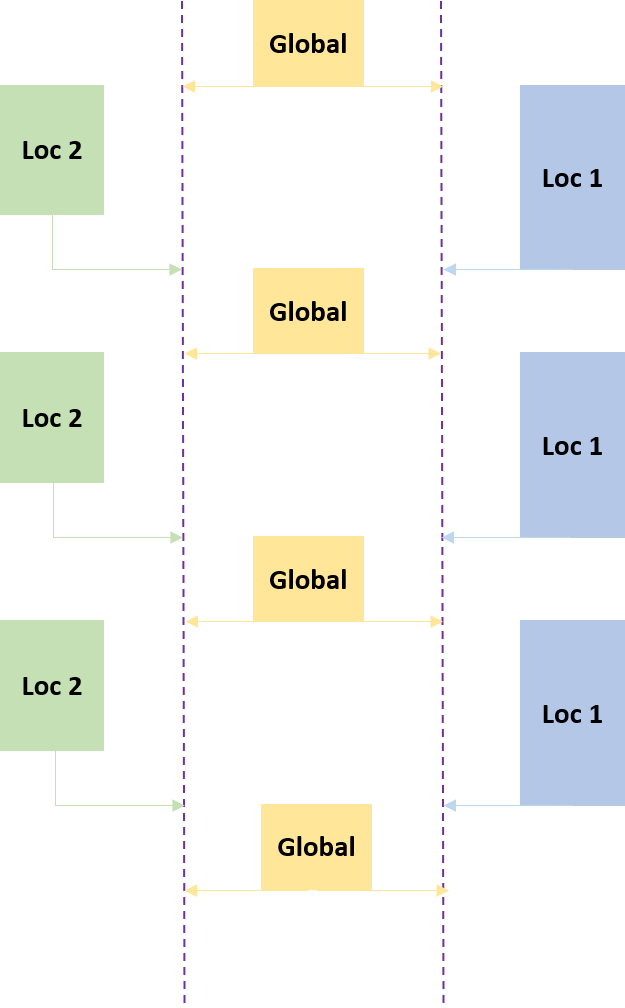}
			\caption{Synchronous model}	\label{Synchronous model}
		\end{subfigure}\hfill
		\begin{subfigure}{.45\textwidth}\centering
			\includegraphics[width=0.61\textwidth]{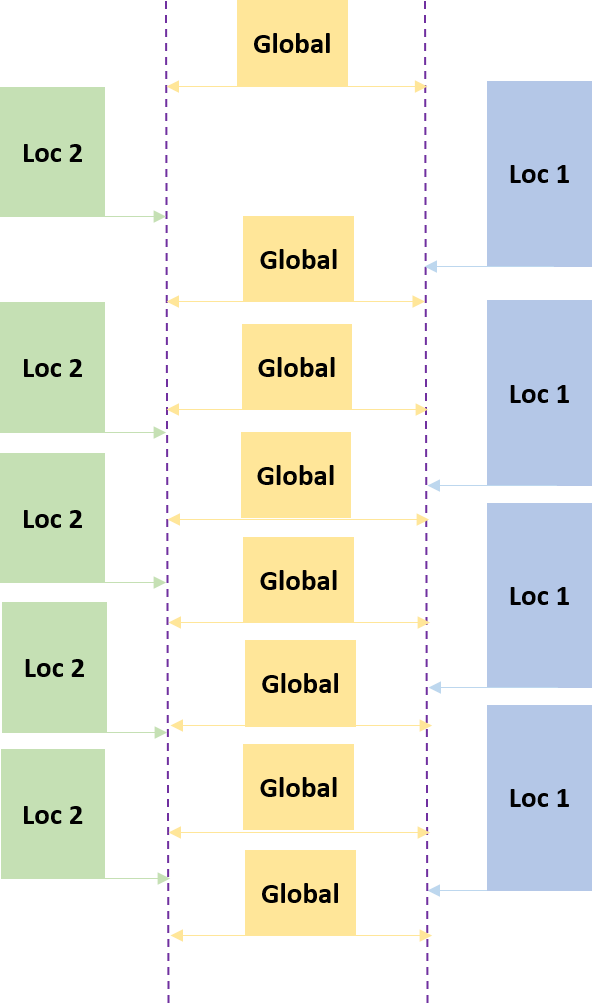}
			\caption{Asynchronous model with wait}\label{Asynchronous with wait}
		\end{subfigure}\hfill\null
		\caption{Time sequence of the synchronous and asynchronous Global-Local coupling.}
	\end{figure}

	Figure~\ref{Synchronous model} presents the time sequence of the classical approach applied to a two-patch case as the one of Figure~\ref{fig:scenario}. The global analysis is carried out in alternation with the local ones and generates waiting and inactivity times on both sides, which seriously affects the method's performance. Problems due to load balancing, communication delays, or machine failures are always related to this synchronization.
	
\section{Asynchronous Coupling}

\subsection{Principle and algorithm}

We investigate the potential of the asynchronous paradigm in order to obtain higher parallel performance than the classical approach. The idea is to launch a computation as soon as some a processor is idling and new input is available.
The asynchronous time sequence is presented in Figure~\ref{Asynchronous with wait}, it is clear that getting rid of synchronization increase the intensity of the computation. 

The iteration is detailed in Algorithm~\ref{alg:staAit}. It can be observed that the Global model is always assembling a residual, which makes it easy to tell convergence, while this point can be troublesome in other methods \cite{Stopping2, Miellou}.

\begin{algorithm2e}[H] \caption{Non-invasive asynchronous iterations}\label{alg:staAit}\DontPrintSemicolon
	Initialization $\bp=0$ and $(\bq\s=0)$ on the \textit{global} model (rank 0)\;
	\If{Rank 0 is available and detects at least one new $\bq\s$} {
		Assemble residual: $\br = - \sum_{s=0}^N\bA\s\bq\s$\; 
		\If{$\|\br\|$ is small enough (initialization excluded)}{break}
		Update $\bp =\bp + \omega\br$\;
		\textit{Global} solves for $\bu\G_{j}$ for given $\bp_{j}$, eq.~\eqref{eq:glo}  \;
		\textit{Global} puts $(\bA\sT\bu\G_A)$ on the patches \;
		If there is a complement, compute $\lam\Gz$  and set $\bq\z=\lam\Gz$
	}
	\If{Subdomain $s>0$ is available and detects new $(\bA\sT\bu\G)$}{
		{Parallel \textit{fine} solve, for $s>0$ obtain $\bu\Fs_j$ and $\lam\Fs_j$ from imposed $\bT\G\bu\G$, eq.~\eqref{eq:loc}}\;
		Subdomain $s>0$ puts  $\bq\s=\bJ\sT\lam\Fs$ on the \textit{Global} model (rank 0)\;
	}
\end{algorithm2e}

The theoretical proof of the convergence of the asynchronous Global-Local iteration can be derived by the framework of paracontractions \cite{paracontractions92}. The main result is that for a given relaxation parameter, the synchronous iteration convergences then so does the asynchronous iteration.

One drawback of the asynchronous iteration is that for now no acceleration strategy is available.

\subsection{Implementation}
Implementing an asynchronous communication protocol has been the subject of several research works, generally based on MPI as in \cite{Gbikpi-Benissan}, where the idea is the use of classical two-sided communication. However, new works based on one-sided communication \cite{Yamazaki.19, Glusa}, also known as MPI-RDMA, have proven the efficiency of these techniques and their adaptation to asynchronous communication.

The performance of these techniques depends on the MPI version used and the network. We tested on several configurations: OPENMPI, INTELMPI, MPICH, and several network architectures like the classic Ethernet and more developed Infiniband or Intel OPA. The influence has been observed in asynchronous with sometimes implicit synchronization imposed by the network or minor performing communication operations depending on the MPI version. 

The general idea of RDMA is to allow access to data on other machines without the need to involve the target machine. We create a part of the memory called a window in which we place the searched data. The other machines will be able to perform operations of type PUT or GET to update this information in the window or to recover it and use it afterward. The idea is thus well adapted to the asynchronous calculation because we are in a procedure where we do not need to stop computing to a Send or a Receive operations.

\begin{figure}[H]
	\centering
	\includegraphics[width=0.5\textwidth]{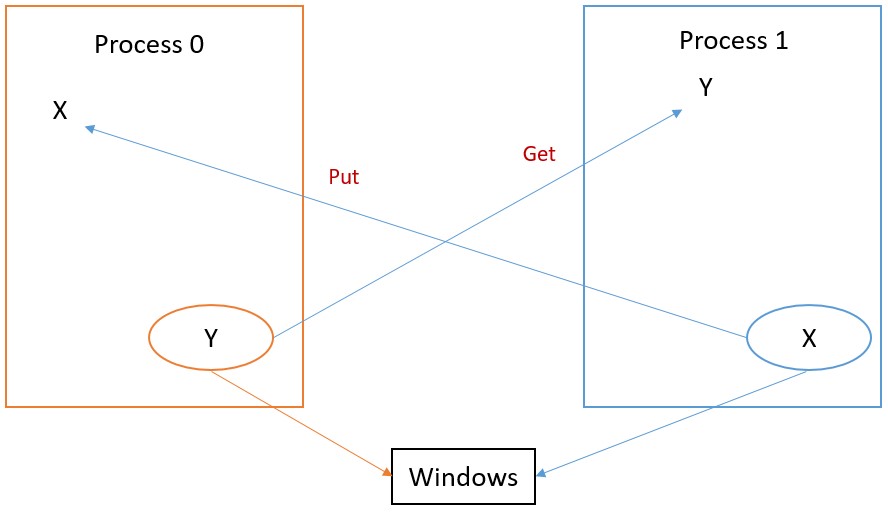}
	\caption{RDMA communications : Put and Get}
	\label{Put_Get}
\end{figure} 

Figure \ref{Put_Get} corresponds to a communication between a processor 0 and processor 1, and this last one applies two operations of communication PUT and GET, on the processor 0, to receive the value X and to send the value Y, these two operations as mentioned before have been made without involving the processor 0.

These communications are usually followed by a synchronization step. We distinguish two types of synchronization in RDMA the active synchronization, where we make  a collective operation to update everyone before going from one iteration to another with the command : \textbf{MPI\_WIN\_Fence()}. The second technique, called passive synchronization, is used in asynchronous. This technique consists of synchronizing each processor without necessarily doing a global synchronization. Each processor opens an epoch with \textbf{MPI\_win\_lock} and produces these \textbf{PUT} and \textbf{GET} operations in this epoch before closing it with \textbf{MPI\_win\_Unlock}. The send completion operations \textbf{MPI\_win\_Flush} follows these operations inside this epoch to ensure that the send is completed.

\section{Numerical results}

Our code is realized in python. It uses several other tools and software like GMSH \cite{GMSH} to generate the geometries and meshes of the studied cases. For the finite element approximation, we use the Getfem library \cite{GetFEM}. For the parallel side, we use the mpi4py library \cite{mpi4py}. 

The study was carried out with the cluster of the LMPS simulation center using several workstations with an ethernet network. These machines are quite heterogeneous with 4 different generation of CPUs :(Intel(R) Xeon(R) CPU E5-1660 v3 (Haswell) @ 3.00GHz,  Intel(R) Xeon(R) CPU E5-2630 v4 (Broadwell) @ 2.20GHz, Intel(R) Xeon(R) Silver 4116 CPU (Skylake) @ 2.10GHz, Intel(R) Xeon(R) W-2255 CPU (Cascade Lake) @ 3.70GHz.

We propose 3 illustrations to analyze the performance of the asynchronous coupling.

\subsection{2D academic cases}
To quickly show the effect and the improvements brought in asynchronous, we study the performances in synchronous and asynchronous of the simple problem presented previously in Figure~\ref{fig:scenario}. The global problem contains 701 nodes, the first patch 381 and the second 379. 
We are interested in a thermal problem (Poisson equation with constant source term) and a linear elasticity problem (under gravitational load). Zero Dirichlet boundary conditions are applied on the bottom side of the Global domain. 

Table~\ref{Thermal_2D} gives the performance for the thermal problem while Table~\ref{Elasticity_2D} deals with the (plane stress isotropic) elasticity problem. We show both computation time and number of iterations, in the asynchronous case, we distinguish between the number of Global iterations and Local iterations for each patch.

\begin{table}[H]
	\centering
	\begin{tabular}{|l|c|c|c|c|}
		\hline
		Models&  \textbf{Synchronous} & \textbf{Aitken} & \textbf{Asynchronous} &  \textbf{Relaxed asynchronous} \\ \hline
		\hline
		{Time}   &  0.22s  & 0.12 s     & 0.3 s & 0.19 s\\ 
		{Iterations} &  23   & 12     &  45[96 - 97] & 29[64 - 65]  \\ \hline
	\end{tabular}
	\caption{Performance for the 2D thermal problem.}
	\label{Thermal_2D}
\end{table}

\begin{table}[H]
	\centering
	\begin{tabular}{|l|c|c|c|c|}
		\hline
		Models&  \textbf{Synchronous} & \textbf{Aitken} & \textbf{Asynchronous} &  \textbf{Relaxed asynchronous} \\ \hline
		\hline
		{Time}   &  0.67s  & 0.3 s     & 0.6 s & 0.52 s\\ 
		{Iterations} &  43   & 16     &  53[112 - 119]  & 48[100 - 107]  \\ \hline
	\end{tabular}
	\caption{Performance for the 2D elasticity problem.}
	\label{Elasticity_2D}
\end{table}

We compare synchronous and asynchronous versions with and without relaxation.
In the synchronous case, relaxation is dynamically adapted using Aitken acceleration. In the asynchronous case, the relaxed iteration corresponds to the best result that we obtained in a trial-and-error campaign to fit the relaxation coefficient. 

The simplicity and well-balanced zones of interest in the problems are against the asynchronous approach. We can see that without relaxation, the asynchronous is faster in the case of linear elasticity. However, (synchronous) Aitken's method provides a good acceleration which cannot be beaten even with carefully chosen relaxation. Anyhow, it is interesting to observe how getting rid of synchronization unleashes computational power: in comparison, much more computations are done with the asynchronous iteration in not as much more time.

\subsection{3D Weak scalability}
\begin{figure}[H]
	\null\hfill
	\begin{subfigure}{0.3\linewidth}\centering
		\includegraphics[scale=0.2]{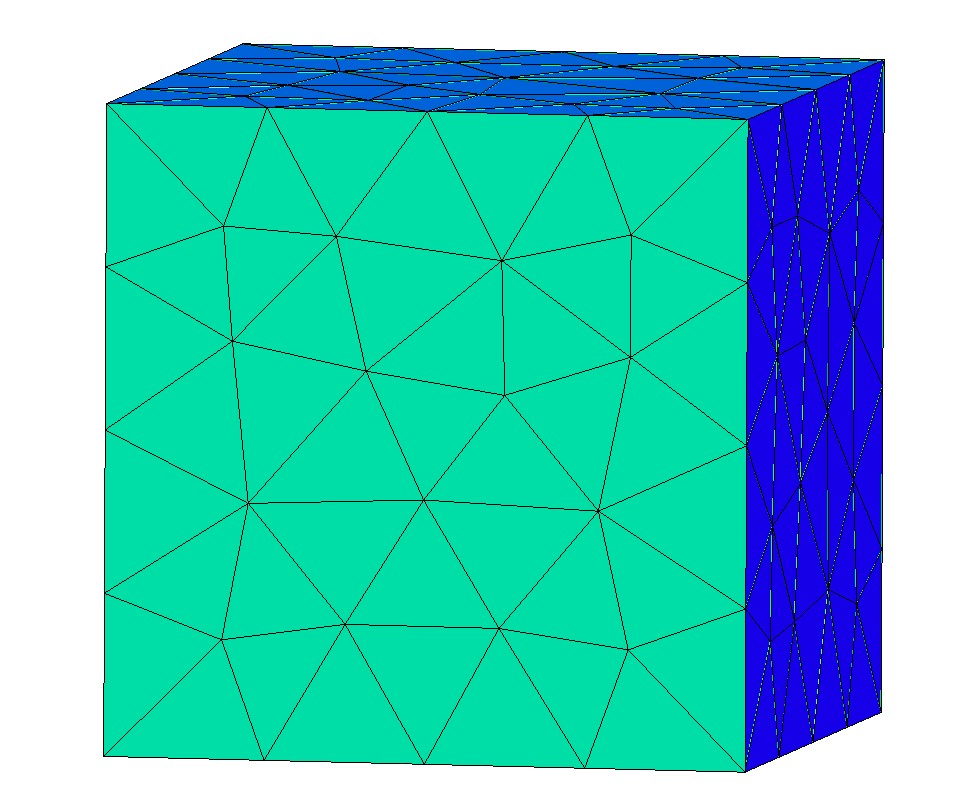}
		\caption{Global $\Omega^{(s)}$}
		\label{Coarse}
	\end{subfigure}\hfill
	\begin{subfigure}{0.3\linewidth}\centering
		\includegraphics[scale=0.2]{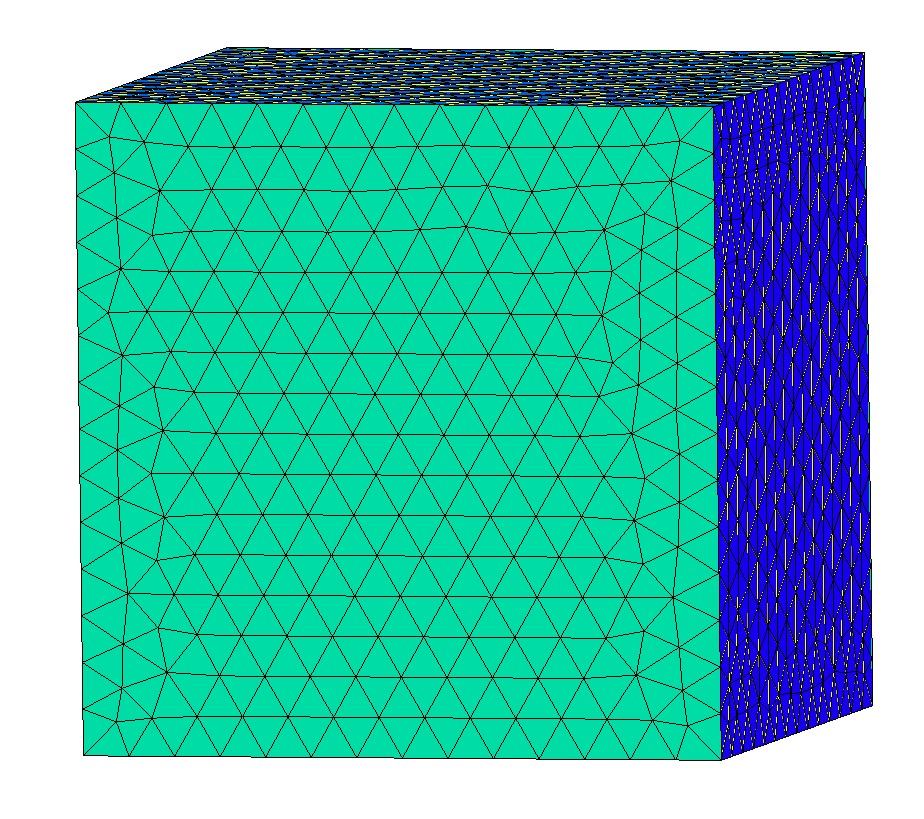}
		\caption{Fin $\Omega^{(s)}$}
		\label{Fin}
	\end{subfigure}\hfill
	\begin{subfigure}{0.3\linewidth}\centering
		\includegraphics[scale=0.3]{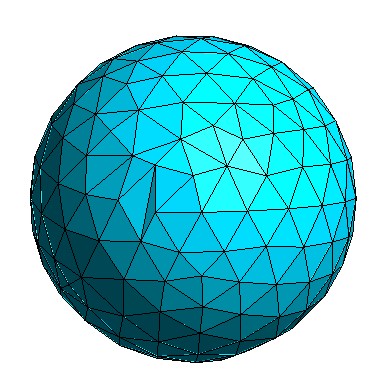}
		\caption{$\Omega^{(s)}_{in}$: Inclusion in Fin $\Omega^{(s)}$}
		\label{inclusion}
	\end{subfigure}\hfill\null
	\caption{Detail of the geometry for the weak scaling test case}
	\label{fig:WSgeom_detail}
\end{figure}

For this study, we generate a simple geometry problem which can easily be extended. The basic pattern is a cube. For the Global model, it is homogeneous and coarsely meshed while the Fine models have a heterogeneous spherical inclusion (the matrix has the same material properties as the Global model) and adapted meshes, see Figure~\ref{fig:WSgeom_detail}. 

\begin{figure}[H]
	\centering
	\includegraphics[scale=0.4]{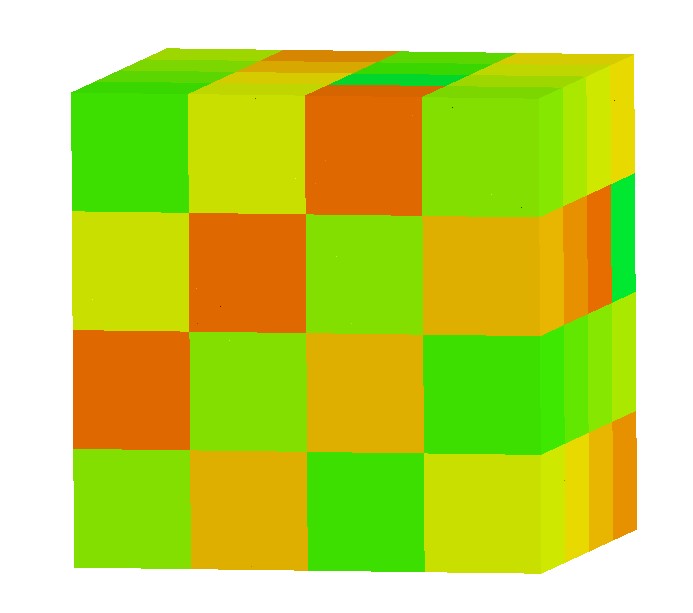}
	\caption{64 subdomains test case for the weak scalability study.}
	\label{64}
\end{figure} 

The idea of the study is to increase the number of subdomains while keeping the same cubic geometry, so we consider the cases with ${n}^3$ subdomains with $ n = 2, ..., 7$. Figure~\ref{64} presents the 64-subdomain global model ($n=4$). Note that the Fine subdomains are perfectly balanced, which is favorable for the synchronous iteration. One face of the Global cube is submitted to zero Dirichlet boundary condition, while a constant source term is applied in the domain.

In Table~\ref{Meshes}, we summarize the number of nodes for each case. The Fine discretization is kept constant, while naturally the size of the Global mesh increases when subdomains are added.

\begin{table}[H]
	\centering
	\begin{tabular}{|l|c|c|c|c|c|c|}
		\hline
		\# of subdomains&  \textbf{8} & \textbf{27} & \textbf{64} &  \textbf{125} &  \textbf{216}&  \textbf{343}\\ \hline
		\hline
		Global   & 233 &   667 & 1449  & 2681  & 4465    &  6903   \\
		Local (1 subdomain) & 1858 &1858     & 1858      & 1858  & 1858  & 1858  \\ \hline
	\end{tabular}
	\caption{Number of nodes in the meshes.}
	\label{Meshes}
\end{table}

\begin{figure}[H]
	\null\hfill
	\begin{subfigure}[b]{.48\textwidth}\centering
		\includegraphics[width=0.95\textwidth]{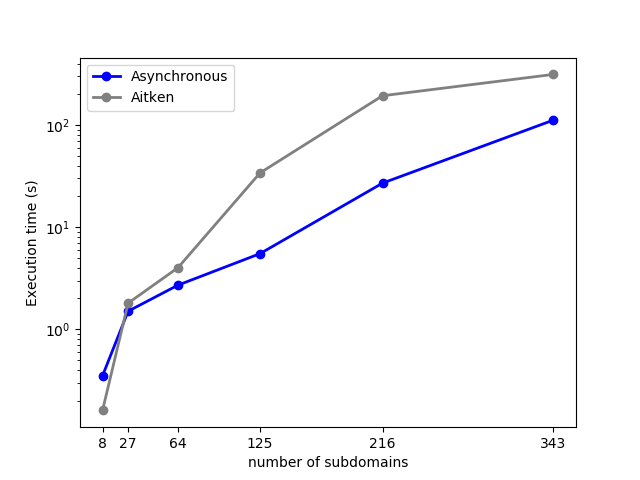}
		\caption{Thermal problem}
		\label{scalability_linear_thermal}
	\end{subfigure} \hfill
	\begin{subfigure}[b]{.48\textwidth}
		\centering
		\includegraphics[width=0.95\textwidth]{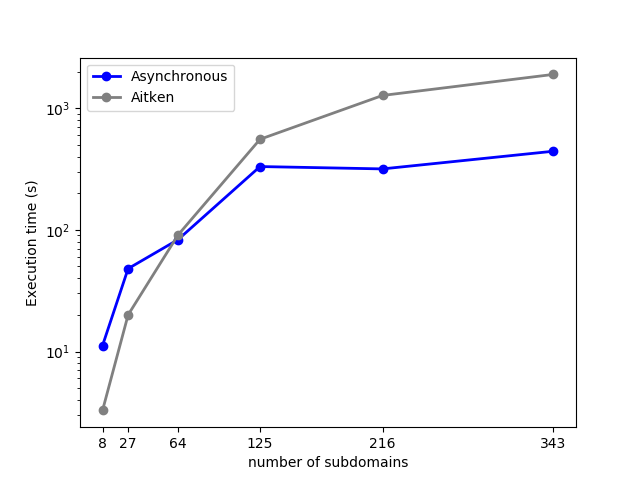}
		\caption{Linear elasticity}
		\label{scalability_linear_elasticity}
	\end{subfigure} \hfill\null
	\caption{Weak scalability study.}
\end{figure}

We first consider a thermal problem where the inclusions are ten time more insulating than the matrix. A comparison of the time to solution for the synchronous (accelerated with Aitken) and asynchronous (with well-chosen relaxation) is presented in  Figure~\ref{scalability_linear_thermal} while Table~\ref{scalability_linear_thermal_table} gives the number of iterations for the Global domain and for the $\max$ and $\min$ number of iterations for the Fine subdomains.

\begin{table}[H]
	\centering
	\small\addtolength{\tabcolsep}{-5pt}
	\begin{tabular}{|l|c|c|c|c|c|c|}
		\hline
		\# subdomains&  \textbf{8} & \textbf{27} & \textbf{64} &  \textbf{125} &  \textbf{216}&  \textbf{343}\\ \hline
		\hline
		Aitken  &  11  & 13    & 12 & 11& 11 & 11  \\
		Asynchronous & 255[32 - 39 ]   & 256[43 - 74]     &  87[49 - 153]  & 65[84 - 207] & 69[276 - 694] & 71[407 - 2902]  \\ \hline
	\end{tabular}
	\caption{Linear thermal (heterogeneity ratio 10) weak scalability study: \# iterations}
	\label{scalability_linear_thermal_table}
\end{table}

We then consider the same geometry for linear elasticity problem. The inclusions are 100 times more flexible than the matrix. Table \ref{scalability_linear_thermal_table} summarizes the number of iterations for each case, while Figure~\ref{scalability_linear_elasticity} compares the CPU time. The behavior of the solver is roughly the same as for the thermal problem, except for small numbers of subdomains for which the synchronous iteration is more efficient.
\medskip

In broad terms, we observe that despite the unfavorable configuration, the asynchronous algorithm is roughly two-time faster than the synchronous one. We can see that this performance is achieved despite the tremendously larger number of iterations.

\begin{table}[H]
	\centering
	\small\addtolength{\tabcolsep}{-5pt}
	\begin{tabular}{|l|c|c|c|c|c|c|}
		\hline
		\# subdomains&  \textbf{8} & \textbf{27} & \textbf{64} &  \textbf{125} &  \textbf{216}&  \textbf{343}\\ \hline
		\hline
		Aitken  &  22 &  21    & 25 & 25 & 26 & 29  \\
		Asynchronous &  2065[78 - 240]   & 1349[102 - 237]     &  372[128 - 475]  & 296[157 - 517] & 295[147 - 514] & 209[175 - 407]  \\ \hline
	\end{tabular}
	\caption{Linear elasticity (heterogeneity ratio 100): \# iterations}
	\label{scalability_linear_elasticityl_table}
\end{table}

\subsection{3D test case}
\begin{figure}[H]
	\centering 
	\begin{minipage}{0.59\linewidth}
		\includegraphics[width=0.99\textwidth]{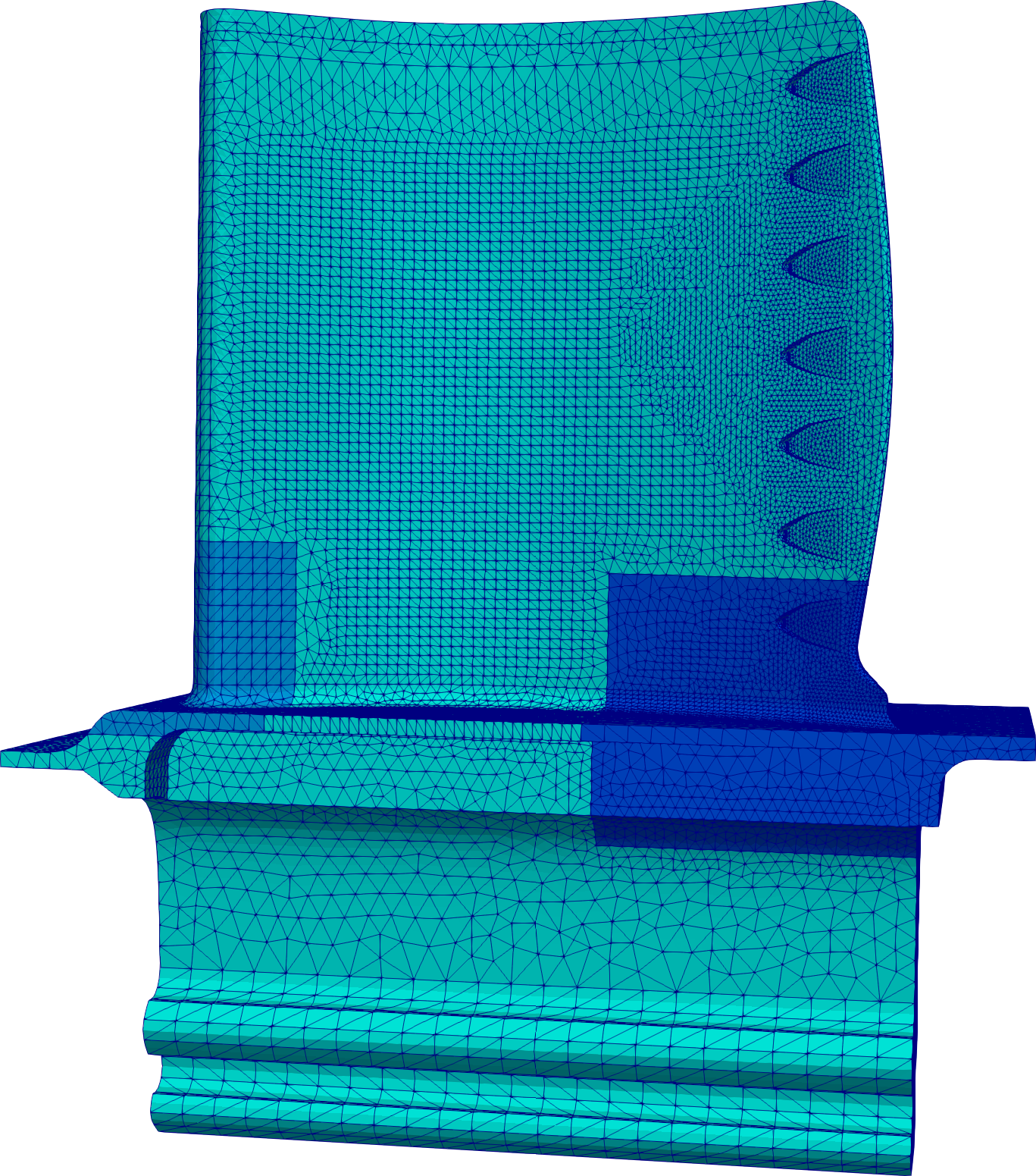}
		\caption{Global problem}
		\label{Global problem aube}
	\end{minipage}
	\hfill
	\begin{minipage}{0.4\linewidth}
		\includegraphics[width=.99\textwidth]{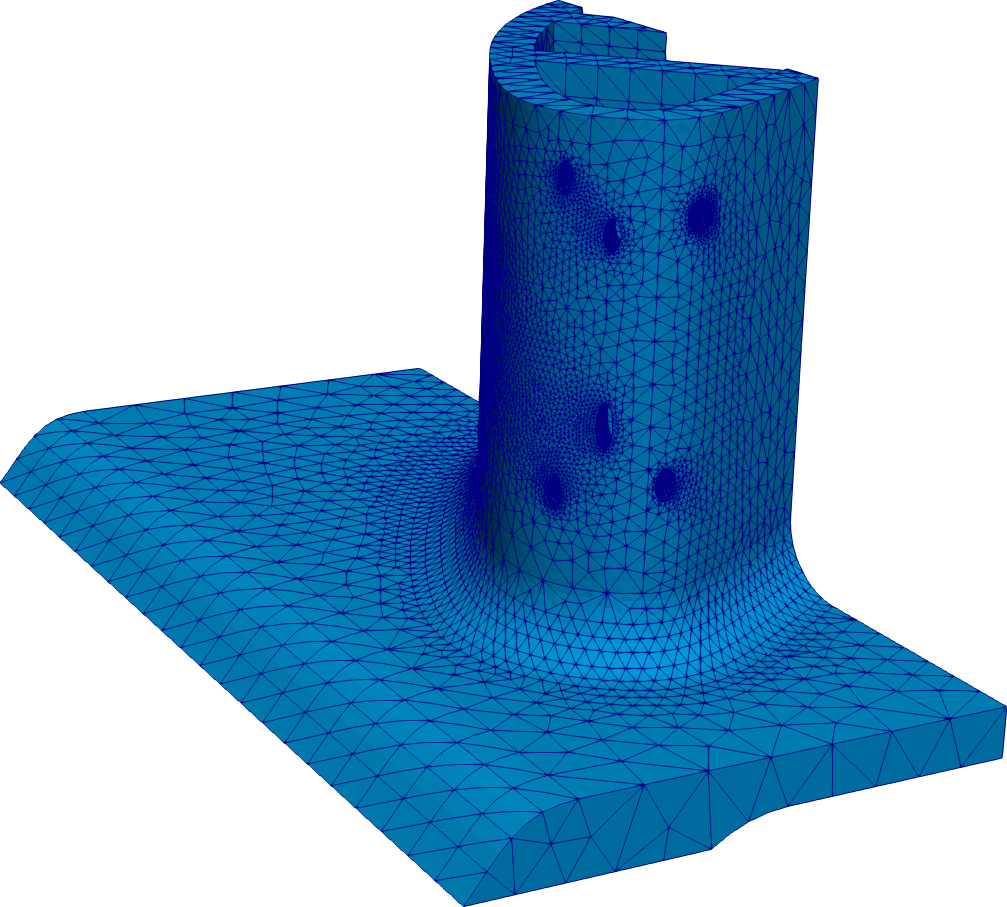}
		
		\includegraphics[width=.99\textwidth]{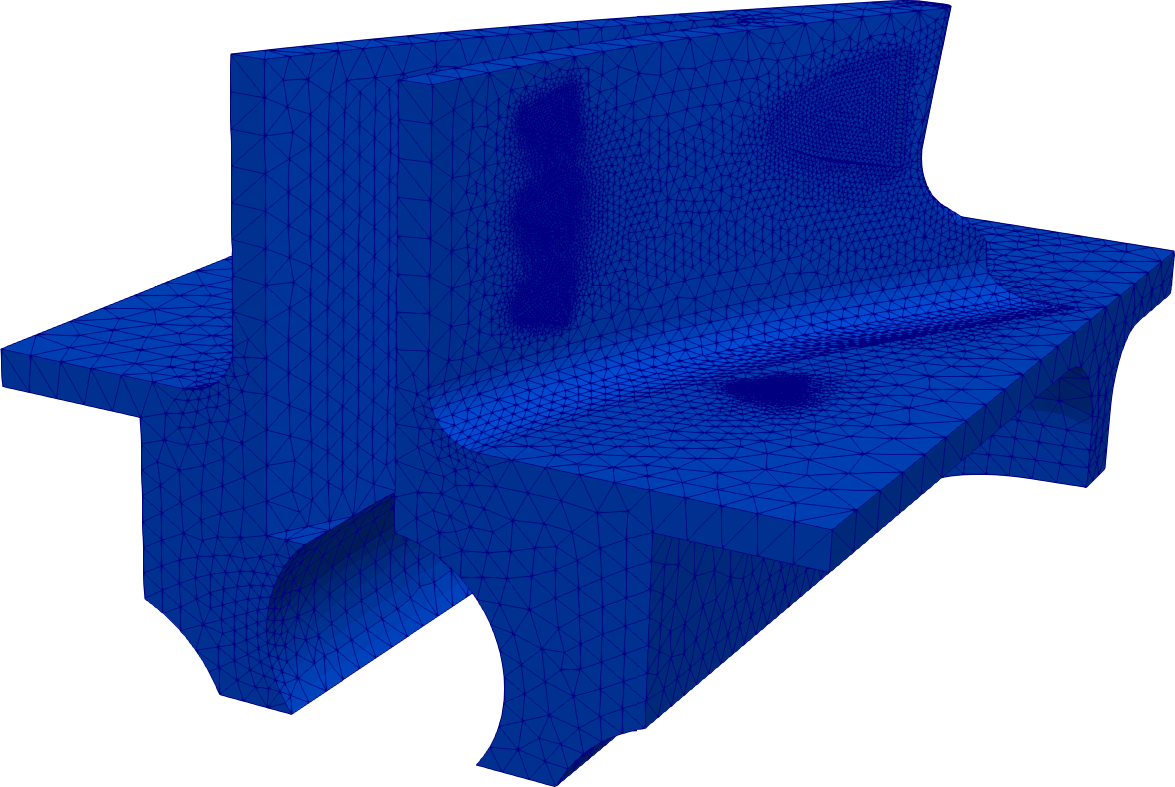}
		\caption{Zones of interest}
		\label{Zones of interest2}
	\end{minipage}
\end{figure}

In this part, we are interested in studying a test case inspired by an industrial problem. The geometry corresponds to the turbine blade of an aircraft engine. The Global model makes use of a simplified geometry which omits cooling micro-perforations. The two zones of interest are two critical regions of the domain where the precise geometry (with the perforations) is taken into account, see Figures~\ref{Global problem aube} and~\ref{Zones of interest2}. The number of nodes of the meshes are given in Table~\ref{Aube3D_mesh}, we can see than one zone of interest is about two times larger than the other one which is roughly of the same size as the Global model.

Details about the actual industrial problem can be found in~\cite{BLANCHARD.2019.1}.  Here, we adopt a simplified version: a thermal problem is considered, with constant source term. In order to make the problem more complex (else very few iterations are needed), we artificially introduce a heterogeneity ratio of 10 (in the conductance coefficient) between the Global and the Fine models.

The parallel analysis is conducted using 3 CPUs: one for the global problem and the other two for each zone of interest. The performance is summed up in Table~\ref{Aube3D}. The asynchronous iteration is about 30\% faster than the accelerated (Aitken) synchronous iteration. Amazingly, we see that the largest subdomains needs fewer iterations in the asynchronous case (20) than in the synchronous case (22); in fact what mattered was having the Global and other subdomain sufficiently converged.

\begin{table}[H]
	\centering
	\begin{tabular}{|l|c|c|c|}
		\hline
		Problem&  \textbf{Global} & \textbf{$1^{st}$ Zone of interest} & \textbf{$2^{nd}$ Zone of interest}\\ \hline
		\hline
		Nodes &  46487 &  40974 &   83900\\ \hline
	\end{tabular}
	\caption{Mesh data}
	\label{Aube3D_mesh}
\end{table}
\begin{table}[H]
	\centering
	\begin{tabular}{|l|c|c|}
		\hline
		Model&  \textbf{Aitken} & \textbf{Asynchronous} \\ \hline
		\hline
		Iterations &  22 &  69[20 - 69]   \\
		Time (s) &  2847.32  &1955.20   \\ \hline
	\end{tabular}
	\caption{Time + Iterations}
	\label{Aube3D}
\end{table}

\section{Conclusion}
In this paper, an asynchronous version of the non-invasive Global-Local coupling has been presented. The MPI-RDMA parallelization techniques with passive synchronization have been used for the programming aspect. The presented results showed that the asynchronous version may allow better performance than the synchronous Aitken accelerator on a heterogeneous cluster.
\subsection*{Acknowledgments}
This work was partly funded by the French National Research Agency as part of project ADOM, under grant number ANR-18-CE46-0008.
\newpage

\bibliography{biblio}
\bibliographystyle{plain}

\end{document}